# Monolithic Photoelectrochemical Device for 19% Direct Water Splitting

Wen-Hui Cheng[1,2]†, Matthias H. Richter[2,3]†, Matthias M. May[4,5,6]*, Jens Ohlmann[7], David Lackner[7], Frank Dimroth[7], Thomas Hannappel[6]*, Harry A. Atwater[1,2]*, Hans-Joachim Lewerenz[2,8]

[1]Department of Applied Physics and Material Science, California Institute of Technology, Pasadena, CA 91125, USA.

[2]Joint Center for Artificial Photosynthesis, California Institute of Technology, Pasadena, CA 91125, USA.

[3] Division of Chemistry and Chemical Engineering, California Institute of Technology, Pasadena, CA 91125, USA.

[4]Department of Chemistry, University of Cambridge, Lensfield Road, CB2 1EW Cambridge, UK

[5]Helmholtz-Zentrum Berlin für Materialien und Energie GmbH, Institute for Solar Fuels, Hahn-Meitner-Platz 1, D-14109 Berlin, Germany.

[6]Department of Physics, Technische Universität Ilmenau, Gustav-Kirchhoff-Str. 5, D-98693 Ilmenau, Germany.

[7]Fraunhofer Institute for Solar Energy Systems ISE, Heidenhofstraße 2, D-79110 Freiburg, Germany.

[8]Division of Engineering and Applied Science, California Institute of Technology, Pasadena, CA 91125, USA.

*Correspondence to:  haa@caltech.edu; thomas.hannappel@tu-ilmenau.de; mm2159@cam.ac.uk

†These authors contributed equally to this work.

**Abstract**

Recent rapid progress in efficiencies for solar water splitting by photoelectrochemical devices has enhanced its prospects to enable storable renewable energy. Efficient solar fuel generators all use tandem photoelectrode structures, and advanced integrated devices incorporate corrosion protection layers as well as heterogeneous catalysts. Realization of near thermodynamic limiting performance requires tailoring the energy band structure of the photoelectrode and also the optical and electronic properties of the surface layers exposed to the electrolyte. Here, we report a monolithic device architecture that exhibits reduced surface reflectivity in conjunction with metallic Rh nanoparticle catalyst layers that minimize parasitic light absorption. Additionally, the anatase $TiO_2$ protection layer on the photocathode creates a favorable internal band alignment for hydrogen evolution. An initial solar-to-hydrogen efficiency of 19.3 % is obtained in acidic electrolyte and an efficiency of 18.5 % is achieved at neutral pH condition (under simulated sunlight).

**Main Text**

Advances in the field of artificial photosynthesis [1] have led to the development of functional prototypes for photoelectrochemical water splitting [2], featuring improved photoelectrode stability through the use of corrosion protection layers [3] and the realization of systems for unassisted water splitting [4-6] in integrated monolithic devices. The requirement for the device operating voltage under illumination to exceed the thermodynamic potential difference for water dissociation of 1.23 V imposes constraints on the energy bandgaps for the photoelectrode absorber layers and their combined operating potential in a series-connected tandem configuration. Several strategies have been followed. Early prototypes used single absorber layers [7-9], but essentially all recent high efficiency devices have used dual-junction tandem structures based on high performance compound semiconductors [4,5], or triple-junction devices comprised of more earth abundant material components [10]. Until now, the highest solar-to-hydrogen (STH) efficiencies were obtained by surface modification of dual-junction III-V compound tandem structures [4-6,11]. Recently, we reported a theoretical analysis of achievable STH efficiency limits that included realistic typical values for parameters of high efficiency materials that account for catalyst performance, optical losses and the absorber external radiative efficiency, which reflects the ratio of radiative to non-radiative recombination [12]. We define under the assumption of perfect light absorber (Shockley Queisser) the theoretical limit for realistic water splitting ($SQ_{water}$). Attaining such limiting efficiencies in practice provides the basis for a renewable fuels technology. The extension of existing fuel infrastructure to hydrogen fuels allows almost direct widespread application in the transportation sector and for power plants [13]. Here, we demonstrate how to achieve efficiencies that approach the theoretical limits for the photoelectrode energy bandgaps employed, by the use of a $TiO_2$ layer that was deposited by atomic layer deposition (ALD). It performs as an interlayer that facilitates near-optimal energy band alignment between the electrolyte and the semiconductor photoelectrode, and controls the optical properties of the front part of the photoelectrochemically active tandem device.

**Integrated photoelecterochemical device and interface energetics**

A dual-junction tandem structure [14,15] with photoelectrode bandgaps tailored towards optimizing water photolysis, is used as a photocathode. With the energy bandgap combination (GaInP/GaInAs with 1.78 eV/1.26 eV, see Supplementary Fig. S1), a theoretical efficiency of 22.8 % under AM 1.5G conditions can be reached, assuming 100 % above bandgap absorption and 100 % external radiative efficiency (ERE) [12]. Although semiconductor photocathodes show enhanced stability compared to photoanodes [16] III-V compound semiconductors are typically unstable with respect to corrosion unless additionally stabilized by transparent, conducting and chemically robust protective layers [17]. To protect the surface from cathodic photocorrosion, we employ a micro-crystalline anatase $TiO_2$ coating (Supplementary Fig. S2) formed by ALD. Rh nanoparticles were employed as hydrogen evolution electrocatalysts and deposited photoelectrochemically under stroboscopic illumination. The resulting device structure is illustrated schematically in Fig. 1a, where the photoelectrode, protective layers, Rh catalyst, GaAs substrate and the sputtered $RuO_x$ counter electrode are depicted (see Methods, Supplementary Fig. S3 for synthesis procedure).

The band alignment spanning between the electrolyte-device interface and the semiconductor photoelectrode is also illustrated. The $TiO_2$ layer performs several functions. In addition to its roles as a protective layer, employing a $TiO_2$ electrolyte interface yields a lower overpotential loss (0.1 eV for the $AlInP/TiO_2/Rh$/electrolyte interface, compared to, e.g., 0.5 eV for the $AlInP/AlInPO_x/Rh$/electrolyte interface without applying $TiO_2$) with regard to hydrogen evolution, which only requires about 50 mV [18]. The higher work function of 4.5 eV for $TiO_2$ leads to a more efficient charge transfer without introducing a barrier for electron injection. The band alignment is inferred from ultraviolet photoelectron spectroscopy (UPS) and UV-vis optical spectroscopy measurements, which are given in Supplementary Figs. S4-S5.

**Water-splitting performance of tandem device**

Fig. 1b shows a summary of STH efficiencies achieved in two-electrode setups for different conditions. The highest efficiency of 19.3 % is observed initially in an acidic perchlorate electrolyte. In a buffered solution, at pH 7, an 18.5 % STH efficiency is reached and, for a device employing an anion exchange membrane to separate the fuel products, an STH of 14.8 % is achieved. In these two-electrode experiments, the STH efficiencies were determined by the operating photocurrent density ($J_{op}$) measured at the potential that the counter electrode assumes for unassisted water splitting ($\eta_{STH} = J_{op} \times \frac{1.23V}{100 mW/cm^2} \times \Lambda$ with $\Lambda$ as the Faradaic efficiency). Details about the efficiency benchmarking of the PEC device under AM1.5G conditions, as well as a discussion of efficiency accuracy, are given in the Methods (see Supplementary Figs. S6-S7). Figure 1c sets our experimental results in the context of recent high efficiency water splitting devices, where limiting efficiencies and corresponding semiconductor bandgaps are indicated. Our device exhibits an efficiency increase of 19 % relative to the current record-holding device [11]. The reported results reach ca. 85 % of the theoretical maximum STH efficiency for the semiconductor bandgaps employed (Supplementary Table S1).

**Approaching theoretical limit with light management**

An analysis of the photocurrent density enhancement in the photoelectrode resulting from modification of the surface optical properties upon deposition of the $TiO_2$ coating is shown in Fig. 2. Fig. 2a indicates an approximately 15 % reduction in absolute reflectivity over the relevant spectral range after applying $TiO_2$, demonstrating its role as an antireflection layer (Supplementary Fig. S8). Upon photoelectrochemical deposition of Rh, the changes in the reflectivity are relatively small, indicating an effective optically almost transparent metallic catalyst layer [19-21]. Fig. 2b shows photocurrent-voltage characteristics for the optimized device, and for the case without a $TiO_2$ antireflection layer between Rh and the photoelectrode surface. 15 % of the overall 22 % increase in the light limiting current density can be related to the improved optical design, and the additional 7 % increase in current density is attributed to the more optimal Rh catalyst nanostructure with respect to particle size and distribution for catalysts supported on $TiO_2$. Details of the catalyst deposition are addressed in Fig. 2c, in which fine control of particle size ranging from 10 nm to 70 nm is achievable by appropriate adjustment of the potential during catalyst deposition. The highest efficiency devices exhibit a catalyst distribution as illustrated in Fig. 2d. The $TiO_2$ microstructure exhibits a leaf-like grain structure, which is then decorated with a uniformly dense layer of ca. 10 nm. Rh nanoparticles of this size are almost fully transparent to the visible light due to the plasmonic absorption peak of Rh which is blue shifted away from the visible light spectrum to the ultraviolet region for particle sizes less than 20 nm [21]. This combination of $TiO_2$ surface micro-topography and Rh particle nanostructure morphology enhances the catalyst/protective layer transparency, enabling near limiting current densities.

**Device stability and production collection**

Stability measurements were conducted under both acidic and neutral electrolyte condition for direct comparison as shown in Fig. 3a. The inset gives the two-electrode photocurrent density vs. time for the initial regime, showing that the photocurrent density for acidic pH decreases, whereas the current density remains stable in neutral pH solutions. At acidic pH, the current density drops from 15 mA/cm$^2$ to 10 mA/cm$^2$ in the first 30 min. For pH 7, the photocurrent density and device performance are stable with the current density fluctuating between 15.5 mA/cm$^2$ and 12 mA/cm$^2$. The spikes indicate the influence of bubble formation and detachment. The dynamics are different due to the change in the reduction mechanism (proton reduction at pH 0, water reduction at pH 7) and the surface tension of the electrolyte. The surface tension of the phosphate buffer is higher than for acidic electrolyte, and exhibits more severe bubble accumulation that reduces the photocurrent density (see Supplementary Fig. S9).

Chronoamperometric tests (at -0.4 V vs. counter electrode) show that the device photocurrent density decreases in acidic electrolyte to small values within 3 h. However, in neutral pH electrolyte, stability over 20 h was demonstrated, with the photocurrent density remaining at 83 % of its initial value. At 12 h into the test, a diurnal cycle was simulated by emersion of the sample in the dark. This step resulted in a substantial current density increase, indicating that the previous reduction in current density did not result in a loss of catalyst or the corrosion of the protective layer (Supplementary Fig. S10).

To confirm near-unity Faradaic efficiency ($\Lambda$) for both proton and water reduction, online gas collection of hydrogen and oxygen was conducted and is shown in Fig. 3b (for details, see Methods). The measured gas volume for oxygen (blue symbols) and hydrogen (red symbols) is overlaid with the expected produced gas volume, as calculated from charge passed through the anode and cathode. In both cases, for pH 0 and pH 7, near unity Faradaic efficiency is confirmed through the agreement between the expected and measured gas volumes. However, whereas the curves for pH 7 stay linear with a constant gas production rate for $H_2/O_2$, as expected from the stability measurements, the curves for pH 0 show a deviation from linearity due to the decreasing photocurrent. From X-ray photoelectron spectroscopy measurements (Supplementary Figs. S11-S12), we conclude that at pH 0, the exposed $TiO_2$ experiences local catalyst detachment and decomposes by chemical etching, degrading its ability to protect the underlying photoelectrode (see Fig. 3c). We find that at neutral pH, the protection layer remains intact and leads to the further prolonged stability, also when compared to the existing benchmarks for considered electrode size.

**Conclusions**

The achievement of high solar-to-hydrogen efficiencies in our devices, relative to limiting efficiencies, derives from the ability to tailor light absorption and electron transport in a protective $TiO_2$ coating that serves as the reactive catalyst support as well as the electrolyte interface. High photocurrent densities require the combination of the antireflection properties of the anatase $TiO_2$ layer, with the use of an optically transparent Rh nanoparticle surface layer. In addition, conduction band alignment through the surface layers across AlInP / AlInPOx / $TiO_2$ / Rh /electrolyte, that promotes the transport of the excess electrons and inhibits voltage drops, is necessary. We demonstrated this achievement in acidic as well as in neutral pH electrolytes. The costs for solar fuel generating systems are more sensitive to the impact of the solar-to-fuel efficiency due to the additional complexity of gas-handling facilities. Therefore, obtaining such high conversion efficiencies for photoelectrochemical hydrogen generation is a prerequisite for widespread application of the technology in the transport sector and for power plants.

**Methods**

**Tandem cell epitaxy.** The dual-junction light absorber ($Ga_{0.41}In_{0.59}P/Ga_{0.89}In_{0.11}As$ with 1.78 eV and 1.26 eV) was grown by metal-organic vapor phase epitaxy in an Aixtron 2800-G4-TM reactor [14,15] on a 4'' p-GaAs wafer using a GaInAs metamorphic step-graded buffer layer to overcome the difference in lattice-constant between the substrate and the solar cell layers. The threading dislocation density after the metamorphic buffer is below $1 \times 10^6$ $cm^{-2}$. In comparison to a previous publication [5] the thickness of the top cell was increased to improve the current matching of the sub cells correlated to the modified spectrum in water [22]. Additionally, switching from n-Ge substrate to p-GaAs removed the requirement of a second tunnel diode below the GaInAs sub-cell.

**Photocathode fabrication.** The native oxide on the back of the GaAs substrate was removed prior to metal ohmic contact deposition by rinsing in acetone; isopropanol; 30 sec $NH_4OH$ (10 %); $H_2O:N_2$ and drying in $N_2$. Immediately afterwards 70 nm Pd, 70 nm Ti and 200 nm Au were deposited by electron beam evaporation followed by rapid thermal annealing at 400 °C for 60 s under $N_2$ atmosphere [23].

Prior to the $TiO_2$ layer deposition, the front GaAs/GaInAs cap layer was removed in a chemical etch bath. The sample was degreased by 15 s rinsing in 2-propanol, 15 s in $H_2O:N_2$ followed by a 60 s etch step in 25 % $NH_4OH$:30 % $H_2O_2$:$H_2O$ (1:1:10), finishing with a 20 s rinse in $H_2O:N_2$ and drying under $N_2$ (Supplementary Fig. S3, step 1). Directly afterwards (a desiccator was used for sample transfer between systems), $TiO_2$ was deposited by atomic layer deposition (ALD) in an Ultratech Fiji F200/G2 ALD system using a titanium tetraisopropoxide (TTiP) precursor (STREM Chemical Inc.) and water as the oxidizer. The deposition temperature was set to 250 °C and a total of 1500 ALD cycles were carried out (Supplementary Fig. S3, step 2).

Note that the edge of the sample has been carefully removed to prevent shunting of the front and back surfaces. Ag paste was applied to attach an ohmic contact to a coiled, tin-plated Cu wire which was then threaded through a glass tube. The sample was encapsulated and sealed to the glass tube using black epoxy (Electrolube ER2162).

The Rh catalyst was photoeletrodeposited (Supplementary Fig. S3, step 3) in an aqueous solution of 0.5 mM Rh(III) chloride trihydrate (99.98%, Sigma Aldrich) + 0.5 M KCl (99.5%, Alfa Aesar) at +0.3 V vs. an SCE reference electrode under pulsed illumination. White light was provided by an Oriel Instruments Solar Simulator using a 1000 W Mercury-Xenon arc lamp. The frequency of the stroboscopic illumination resulted from the optical chopper frequency and the double structure of the chopper wheel. The resulting current profile is shown as an inset in Fig. 2c.

**Counter electrode fabrication.** Counter electrodes were prepared by sputtering ruthenium for 60 min on titanium foil (0.125 mm, 98 %, Sigma Aldrich) using an AJA sputtering system with a forward RF power of 200 W, 5 mTorr Ar atmosphere and a base pressure of $2x10^{-8}$ mTorr. The as prepared electrodes were cut into 1 $cm^2$ pieces, attached with Ag paste to a tin-plated Cu wire which was then threaded through a glass tube. The counter electrode sample was encapsulated and sealed to the glass tube using black epoxy (Electrolube ER2162).

**Photoelectrochemical measurements.** All photoelectrochemical measurements were performed using Biologic SP-200 potentiostats. 1 M $HClO_4$ was used as the electrolyte for pH 0 and 0.5 M $KH_2PO_4/K_2HPO_4$ phosphate buffer for pH 7. All electrolytes were purged with $N_2$ (4N) for minimum 1 h before usage. A saturated calomel electrode (SCE) was used as the reference electrode for three-electrode measurements. Glass cells with a quartz window where used as the vessel for the experiments allowing them to be easily cleaned in Aqua Regia. To avoid internal reflections in the cell, a black mask was directly attached in front of the quartz window so that only the sample itself was illuminated. The tandem device was positioned 10 mm away from the quartz window with the counter electrode being placed in close vicinity at the back of the working electrode. J-V measurements were performed with a scan velocity of 50 mV/s.

Stability and efficiency tests in two electrode configuration were carried out using a calibrated AAA grade AM1.5G solar spectrum provide by an ABET Technologies Sun 3000

Solar Simulator (Supplementary Fig. S6). The light intensity was set to 100 mW/cm$^2$ using a calibrated silicon reference solar cell.

**Gas collection.** Hydrogen and oxygen gas collection were performed using an eudiometric gas collection setup. A SELEMINON ion exchange membrane was utilized to separate the cathode and anode chamber. Electrolytes were purged with ultrapure N$_2$ (4N) and the anode side was saturated in addition with O$_2$. Each side was sealed against the ambient but connected via a short thin tubing to an inverted water filled buret (purged and saturated). The change in pressure in each buret upon H$_2$ and O$_2$ gas collection due to photoelectrochemical water splitting in the PEC cell was monitored by pressure transducers (EXTECH HD755). The change in pressure over time was then converted to a gas volume under consideration of the reduced pressure in the inverted buret. The expected produced volume of hydrogen and oxygen gas for the cathodic and anodic reaction was calculated by the transferred electrical charge as measured by the potentiostat.

**Assessment of solar-to-hydrogen efficiency measurement.** In order to consider the influence of the spectral mismatch of the irradiance between our solar simulator and the AM1.5G spectrum, a spectral correction factor (SCF) was calculated (see Supplementary Fig. S7). It is based on the relative EQE of the device under test (Fig. S1), the irradiance of the solar simulator {$I_{meas}(\lambda)$} (Supplementary Fig. S6), and the AM1.5G reference spectrum {$I_{AM1.5G}(\lambda)$}. The influence of the water filter {$F_{water}(\lambda)$} on the spectra was considered for the calculations. The index j denotes to the individual sub cell.

$$J_{AM1.5G} = J_{meas} \cdot \frac{\min \int_{280nm}^{1200nm} I_{AM1.5G}(\lambda) \cdot F_{water}(\lambda) \cdot EQE_{j,device}(\lambda) d\lambda}{\min \int_{280nm}^{1200nm} I_{meas}(\lambda) \cdot F_{water}(\lambda) \cdot EQE_{j,device}(\lambda) d\lambda} = J_{meas} \cdot SCF \quad (1)$$

For illumination under AM1.5G conditions the reference spectrum was taken from the Renewable Resource Data Center (RReDC) of the National Renewable Energy Laboratory (NREL) [24].

To correct for the not completely parallel light beam illumination in the solar simulator that results in focusing of the light by the quartz window, the beam divergence in each axis was experimentally determined and a concentration ratio (CR) was calculated (Supplementary Fig. S7). The corrected photocurrent is given by $J_0 = \frac{J_{meas}}{CR}$.

The exposed electrode surface area could be precisely determined using an optical scanner and the open source software ImageJ. The area of semi-transparent epoxy at the boundary of the sample was included as well to consider the full photocurrent generating area [22]. In this study the electrodes had different areas of 0.1 – 0.3 cm$^2$.

The total correction factor for each sample is then given by $J_{AM1.5G} = J_{meas} \cdot SCF/CR$, e.g. for the 19.3 % efficient cell reported in Fig. 1b, the values are SCF = 1.024 and CR = 1.028.

**Optical and surface analyses.** Optical measurements were performed to obtain reflectivity spectra for different surface layer stackings in air. A Cary 5000 UV/vis/NIR with integrating sphere that include diffuse reflectivity measurement was used.

For surface topography studies, a Bruker Dimension Icon AFM in Peakforce mode was used. Scanning electron microscopy images were obtained with a FEI Nova NanoSEM 450 microscope.

**References:**

1. Walter, M. G. *et al.* Solar Water Splitting Cells. *Chem. Rev.* **110,** 6446–6473 (2010).
2. Xiang, C. *et al.* Modeling, Simulation, and Implementation of Solar-Driven Water-Splitting Devices. *Angew. Chem. Int. Edit.* **55,** 12974–12988 (2016).
3. Lichterman, M. F. *et al.* Protection of inorganic semiconductors for sustained, efficient photoelectrochemical water oxidation. *Catal. Today* **262,** 11–23 (2016).
4. Khaselev, O. & Turner, J. A. A monolithic photovoltaic-photoelectrochemical device for hydrogen production via water splitting. *Science* **280,** 425–427 (1998).
5. May, M. M., Lewerenz, H. J., Lackner, D., Dimroth, F. & Hannappel, T. Efficient direct solar-to-hydrogen conversion by in situ interface transformation of a tandem structure. *Nat. Commun.* **6,** 8286 (2015).
6. Verlage, E. *et al.* A monolithically integrated, intrinsically safe, 10% efficient, solar-driven water-splitting system based on active, stable earth-abundant electrocatalysts in conjunction with tandem III–V light absorbers protected by amorphous $TiO_2$ films. *Energ. Environ. Sci.* **8,** 3166–3172 (2015).
7. Fujishima, A. Electrochemical Photolysis of Water at a Semiconductor Electrode. *Nature* **238,** 37–38 (1972).
8. Krasnovsky, A. A. & Nikandrov, V. V. The photobiocatalytic system: Inorganic semiconductors coupled to bacterial cells. *FEBS Letters* **219,** 93–96 (1987).
9. Kudo, A., Ueda, K., Kato, H. & Mikami, I. Photocatalytic $O_2$ evolution under visible light irradiation on $BiVO_4$ in aqueous $AgNO_3$ solution. *Catalysis Letters* **53,** 229–230 (1998).
10. Abdi, F. F. *et al.* Efficient solar water splitting by enhanced charge separation in a bismuth vanadate-silicon tandem photoelectrode. *Nat. Commun.* **4,** 2195 (2013).
11. Young, J. L. *et al.* Direct solar-to-hydrogen conversion via inverted metamorphic multi-junction semiconductor architectures. *Nature Energy* **2,** 17028 (2017).
12. Fountaine, K. T., Lewerenz, H. J. & Atwater, H. A. Efficiency limits for photoelectrochemical water-splitting. *Nat. Commun.* **7,** 13706 (2016).
13. Sathre, R. *et al.* Life-cycle net energy assessment of large-scale hydrogen production via photoelectrochemical water splitting. *Energy Environ. Sci.* **7,** 3264–3278 (2014).
14. Dimroth, F., Beckert, R., Meusel, M., Schubert, U. & Bett, A. W. Metamorphic $GayIn_{1-}$

$_y$P/Ga$_{1-x}$In$_x$As tandem solar cells for space and for terrestrial concentrator applications at C > 1000 suns. *Prog. Photovoltaics* **9,** 165–178 (2001).

15. Ohlmann, J. *et al.* Recent development in direct generation of hydrogen using multi-junction solar cells. *AIP Conf. Proc.* **1766,** 080004 (2016).

16. Heller, A., Miller, B., Lewerenz, H. J. & Bachmann, K. J. An efficient photocathode for semiconductor liquid junction cells: 9.4% solar conversion efficiency with p-InP/VCl$_3$-VCl$_2$-HCl/C. *J. Am. Chem. Soc.* **102,** 6555–6556 (1980).

17. Hu, S. *et al.* Amorphous TiO$_2$ coatings stabilize Si, GaAs, and GaP photoanodes for efficient water oxidation. *Science* **344,** 1005–1009 (2014).

18. Bolton, J. R., Strickler, S. J. & Connolly, J. S. Limiting and realizable efficiencies of solar photolysis of water. *Nature* **316,** 495–500 (1985).

19. Porter, J. D., Heller, A. & Aspnes, D. E. Experiment and theory of 'transparent' metal films. *Nature* **313,** 664–666 (1985).

20. Degani, Y. *et al.* 'Transparent' metals: preparation and characterization of light-transmitting palladium, rhodium, and rhenium films. *J. Electroanal. Chem.* **228,** 167–178 (1987).

21. Sanz, J. M. *et al.* UV Plasmonic Behavior of Various Metal Nanoparticles in the Near- and Far-Field Regimes: Geometry and Substrate Effects. *J. Phys. Chem. C* **117,** 19606–19615 (2013).

22. May, M. M. *et al.* On the benchmarking of multi-junction photoelectrochemical fuel generating devices. *Sustainable Energy Fuels* **1,** 492–503 (2017).

23. Chor, E. F., Zhang, D., Gong, H., Chong, W. K. & Ong, S. Y. Electrical characterization, metallurgical investigation, and thermal stability studies of (Pd, Ti, Au)-based ohmic contacts. *J. Appl. Phys.* **87,** 2437 (2000).

24. Solar Spectral Irradiance: ASTM G-173. *rredc.nrel.gov* Available at: http://rredc.nrel.gov/solar/spectra/am1.5/ASTMG173/ASTMG173.html. (Accessed: 29 March 2017)

25. Pourbaix, M. *Atlas of electrochemical equilibria in aqueous solutions*. (Pergamon, 1966).

**Acknowledgments:** This work was supported through the Office of Science of the U.S. Department of Energy (DOE) under award no. DE SC0004993 to the Joint Center for Artificial Photosynthesis, a DOE Energy Innovation Hub. Research was in part carried out at the Molecular Materials Research Center of the Beckman Institute of the California Institute of Technology. The work on tandem absorbers was funded by the German Federal Ministry of Education and research (BMBF) under the contract number FKZ 03F0432A (HyCon). M.M.M acknowledges funding from the fellowship programme of the German National Academy of Sciences Leopoldina, grant LPDS 2015-09. The authors acknowledge Katherine T. Fountaine for the calculation of theoretical photocurrent efficiencies of 2J PEC devices.

**Author Contributions:** T.H., H.J.L, M.M.M., W.H.C., M.H.R. and H.A.A. conceived of the experimental study. W.H.C. and M.H.R. executed the experiments and did the data analysis. J.O., D.L. and F.D. prepared the tandem absorber. W.H.C., M.H.R. and H.J.L. wrote the paper and all authors commented on the manuscript.

**Competing Financial Interests statement:** The authors declare no competing financial interests.

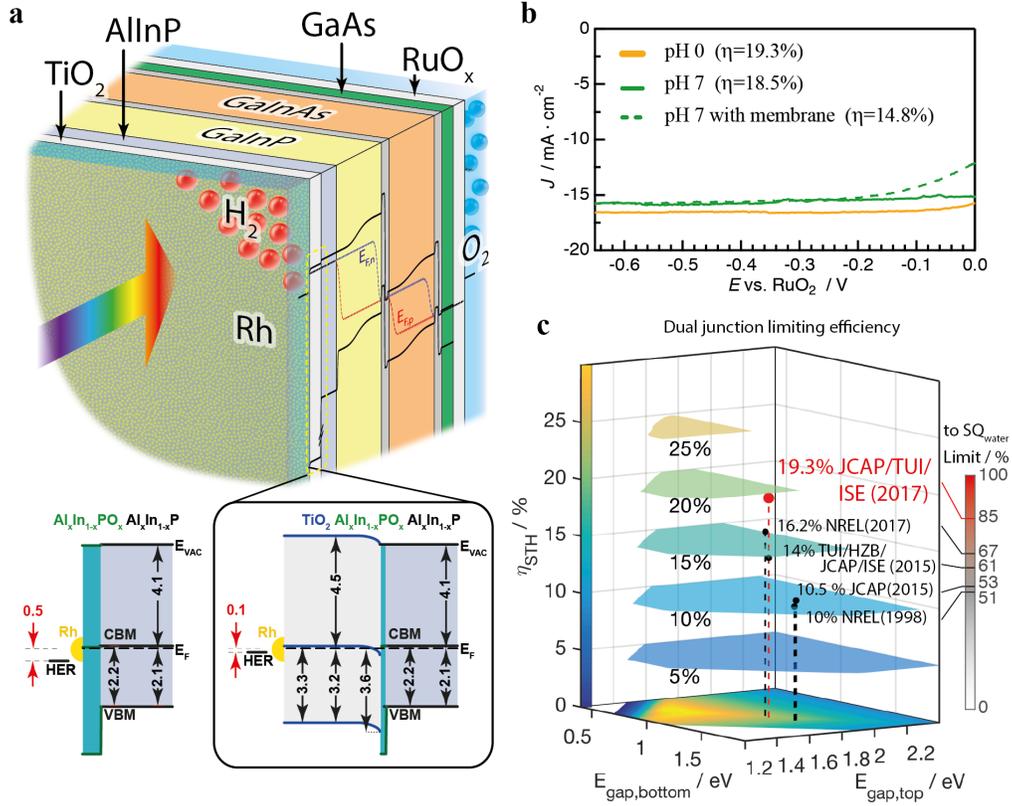

**Fig. 1**. **Efficient integrated photoelectrochemical device for solar fuel production.** (**a**) Illustration of the photoelectrochemical water splitting device structure. Band alignment at the operation point is depicted on the side. At bottom left, the surface band alignment of the electrolyte interface layers with and without $TiO_2$ are shown as a comparison. (**b**) J-V measurements performed under simulated AM1.5G condition for two pH conditions. (**c**) Calculations of the STH efficiency of a dual-junction photoelectrochemical device for 100 % absorption above the bandgap, 100 % ERE, catalytic exchange current density $J_{0,cat} = \{1, 1E-3\}$ mA·cm$^{-2}$, $R_S = 0$ Ω and $R_{SH} = \infty$ Ω including the reported efficiencies and the energy gap pairing. The bar chart on the right indicates the achieved efficiency with respect to the respective theoretical limit (SQ$_{water}$).

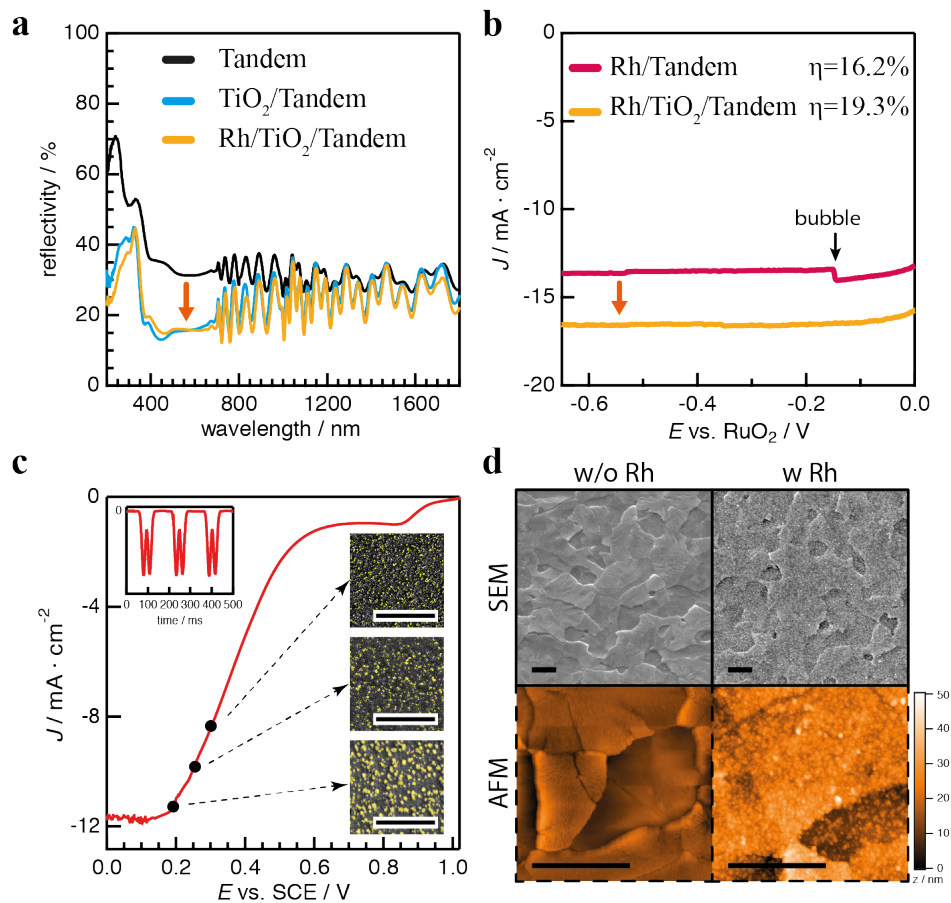

**Fig. 2. Optical and structural surface characterization.** (**a**) Reflectivity measurements on the dual-junction light absorber (black curve), with additional deposited TiO$_2$ coating (blue curve) and with photoelectrochemically deposited Rh catalyst nanoparticles (orange curve). (**b**) Improvement of the photocurrent upon incorporation of a crystalline TiO$_2$ coating by ALD and with an optically optimized catalyst layer. (**c**) Catalyst deposition control with respect to effective potential and current. The upper left inset depicts pulsed illumination. The scale bar is 2 μm. (**d**) SEM images and AFM microtopographs of the dual-junction PEC device with TiO$_2$ coating with and without Rh catalyst nanoparticles. The scale bar is 500 nm. The AFM images are scaled to the same 50 nm z-axis dynamic range.

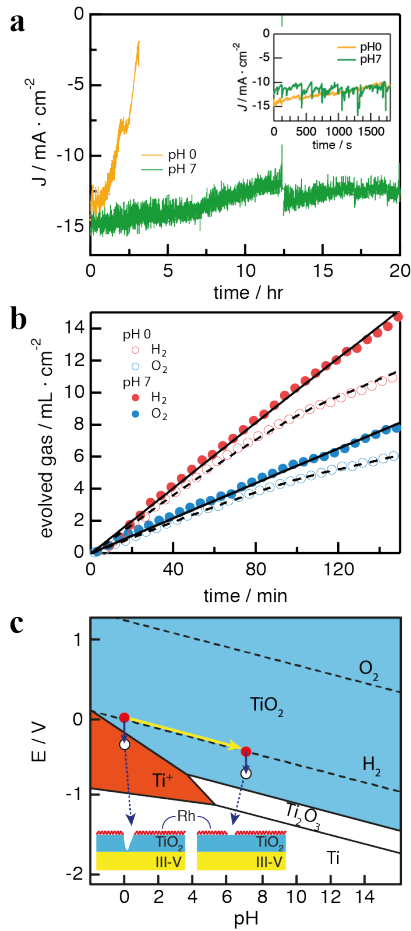

**Fig. 3. Stability and efficiency assessment.** (**a**) Stability measurements at -0.4 V vs. RuO$_x$ counter electrode for acidic and neutral pH. The inset shows the stability data for unassisted water splitting. (**b**) Corresponding gas collection data. Solid and open symbols represent eudiometric gas collection measurements for pH 7 and pH 0 with red and blue colors indicating H$_2$ and O$_2$; solid and dashed lines represent the gas volume calculated from passed charge for pH 7 and pH 0. The production rate exhibits a ratio of 2:1 for H$_2$ to O$_2$. (**c**) Potential-pH equilibrium diagram for the system titanium-water system at 25 °C, taken ref. [25]. For pH 0, the stable region is small. Upon 0.15 V overpotential to hydrogen evolution, corrosion sets in which ultimately leads to the degradation of the device and its efficiency.

# Supplementary Materials for

## Monolithic Photoelectrochemical Device for 19% Direct Water Splitting

Wen-Hui Cheng†, Matthias H. Richter†, Matthias M. May*, Jens Ohlmann, David Lackner, Frank Dimroth,

Thomas Hannappel*, Harry A. Atwater*, Hans-Joachim Lewerenz

correspondence to: haa@caltech.edu; thomas.hannappel@tu-ilmenau.de; mm2159@cam.ac.uk

**This PDF file includes:**

Supplementary Text

Figs. S1 to S12

Tables S1

References (26)

**Supplementary Text**

**External quantum efficiency measurement.** External Quantum Efficiency (EQE) measurements were performed on fully processed tandems. To avoid hydrogen evolution and $H_2$ bubble formation during EQE measurements, a 50 mM methyl viologen hydrate (98%, ACROS Organics), dissolved in ultrapure water, was used as the electrolyte. For continuous light biasing of each individual tandem sub-cell during EQE measurements of the complementary sub-cell, a 780 nm high-power LED (Thorlabs M780L2) was used to bias the bottom cell and a 455 nm high-power LED (Thorlabs M455L2) was used to bias the top cell. Monochromatic illumination was delivered by an Oriel Solar Simulator with a 150 W Mercury-Xenon arc lamp attached to a Newport monochromator (1200 lines/mm). The monochromatic light was chopped at 10.241 Hz. The modulated photocurrent was amplified by an SRS model SR570 low noise current preamplifier. The current preamplifier was also used to supply a -1 V bias to the tandem working electrode. A coiled Pt wire was used as the counter electrode for this two-electrode measurement.

The output from the preamplifier was then measured by a SRS model SR830 lock-in amplifier which was phase locked to the frequency of the optical chopper yielding the photocurrent for the individual sub-cell $J_{top/bottom}(\lambda)$.

To measure the absolute light intensity (W·nm$^{-1}$·cm$^{-2}$) as delivered by the monochromator, a certified calibrated silicon diode (biased at -1 V) was positioned in the light path inside the photoelectrochemical cell filled with the electrolyte (to exclude effects of the electrolyte and quartz window on the measured light intensity) and the photocurrent density was measured (the LED's for light biasing of the tandem were switched off during this reference scan). The photocurrent density could then be converted to the light intensity $I(\lambda)$ by the known spectral response of the silicon diode.

The EQE for each sub-cell is then given by equation (2).

$$EQE_{top/bottom}(\lambda) = \frac{R_{top/bottom}(\lambda)}{\lambda} \cdot \frac{hc}{e} = \frac{J_{top/bottom}(\lambda)}{I(\lambda)} \cdot \frac{1}{\lambda} \cdot \frac{hc}{e} \quad (2)$$

$J_{top/bottom}(\lambda)$ is the photocurrent density of the corresponding sub-cell in $A \cdot nm^{-1} \cdot cm^{-2}$, $I(\lambda)$ the light intensity delivered by the monochromator in $W \cdot nm^{-1} \cdot cm^{-2}$, $\lambda$ is the wavelength in nm, h is the Planck constant, c is the speed of light in a vacuum, and e is the elementary charge. $R_{top/bottom}$ is the spectral response for each sub-cell.

**Photoelectron spectroscopy.** XPS measurements were performed using a Kratos Axis Ultra and Surface Science M-Probe system with a base pressure of < $1 \times 10^{-9}$ mTorr. A monochromatic AlKa (hK = 1486.69 eV) source with a power of 150 W was used for all measurements. He I ultra violet photoelectron spectroscopy (UPS) was performed on the Kratos Axis Ultra system using a Helium gas discharge lamp.

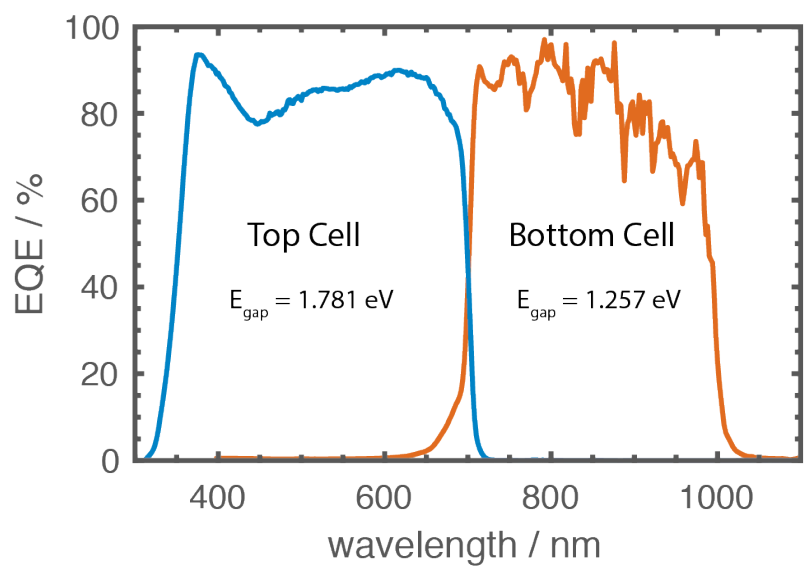

**Fig. S1.**

Relative EQE of fully processed PEC tandem device. Measurements were performed in 50 mM methyl viologen. The bias light was 780 nm and 455 nm for bottom and top sub-cell, respectively.

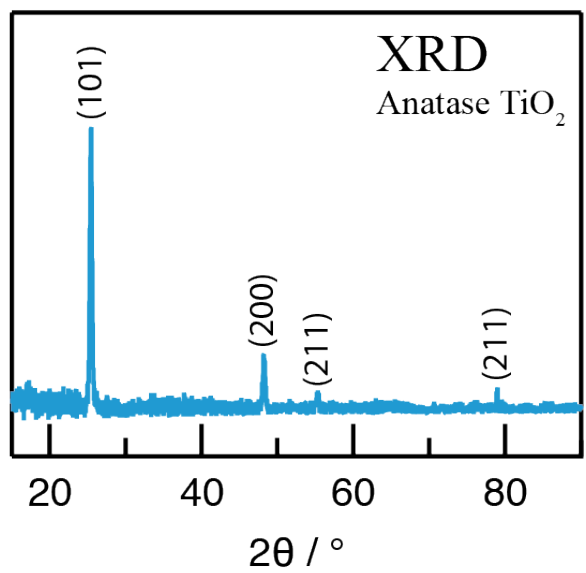

**Fig. S2**

X-ray diffraction data showing anatase phase for TTiP TiO$_2$.

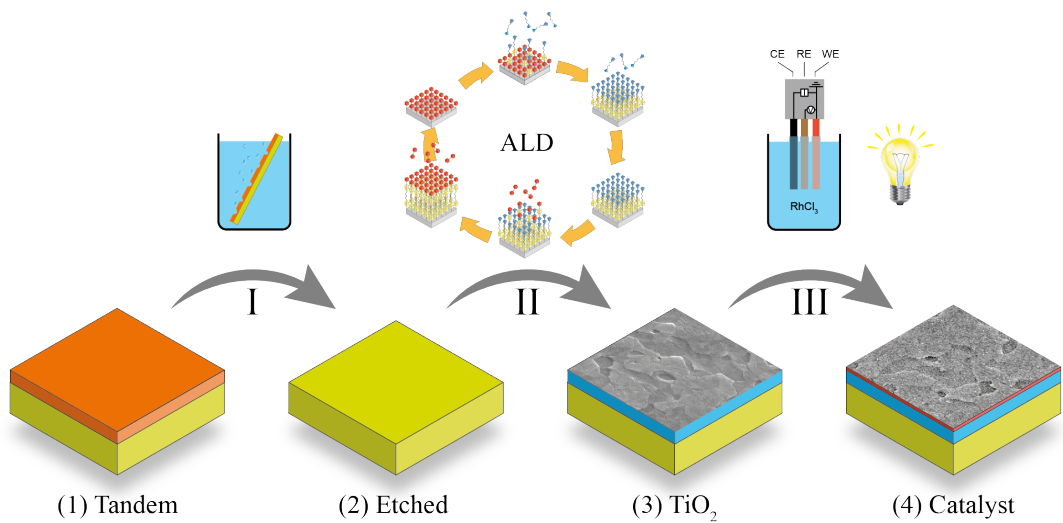

**Fig. S3**

Process flow for preparing the PEC device: (I) Chemical etching of the GaAs/GaInAs cap layer stopping at the AlInP window layer. (II) Deposition of the $TiO_2$ protection and antireflection coating with ALD. (III) Photoelectrochemical deposition of a closed layer of Rh nanoparticles onto the tandem.

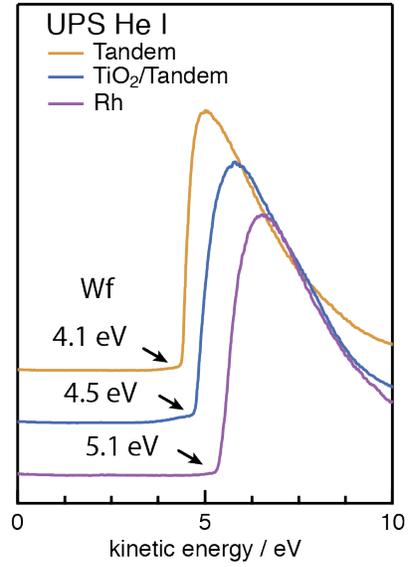

**Fig. S4**

Workfunction measurements by UPS for the tandem, for $TiO_2$ on the tandem and for Rh metal. The increase of workfunction from 4.1 eV to 4.5 eV was observed after applying $TiO_2$ protection layer on tandem.

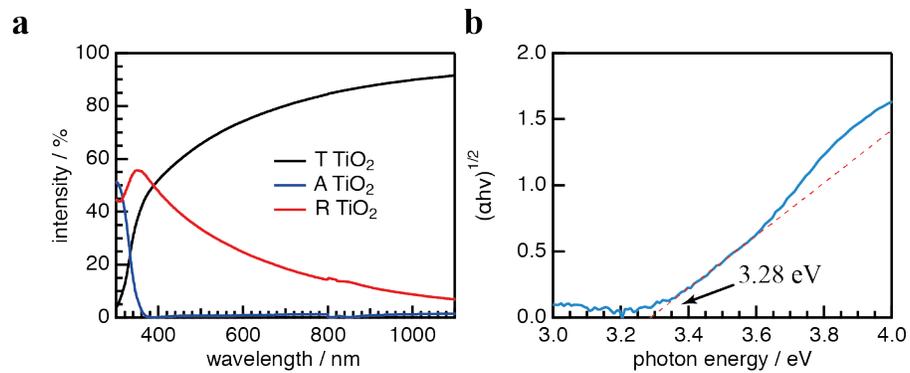

**Fig. S5**

(**a**) Optical properties (A: absorption, T: transmission, R: reflection) of TiO$_2$ (TTiP ALD). (**b**) Tauc plot of ALD grown TiO$_2$. The intersection with the horizontal axis indicates an indirect optical gap of 3.28 eV.

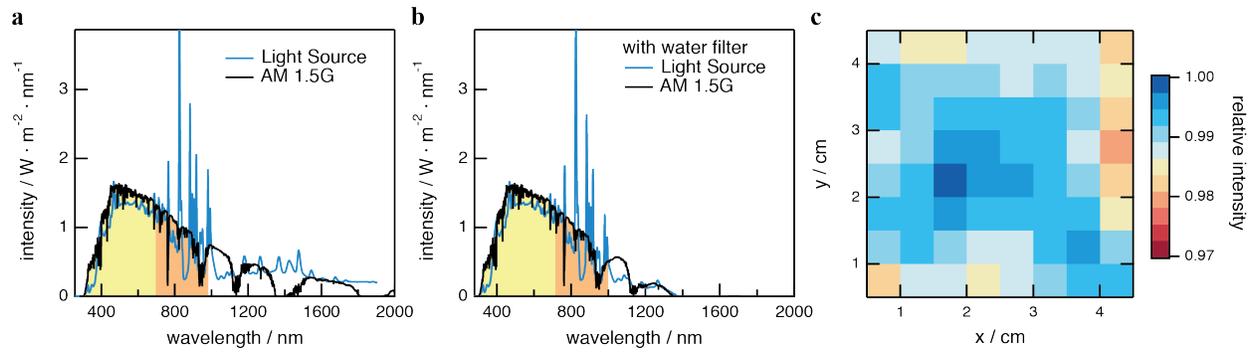

**Fig. S6**

(**a**) Light spectrum of the solar simulator (ABET Sun 3000 Solar Simulator) and AM1.5G spectrum. (**b**) Light spectrum of the solar simulator and AM1.5G with water filter. (**c**) Uniformity map of the illumination area. The band gaps of the dual-junction light absorber are indicated in (**a**) and (**b**).

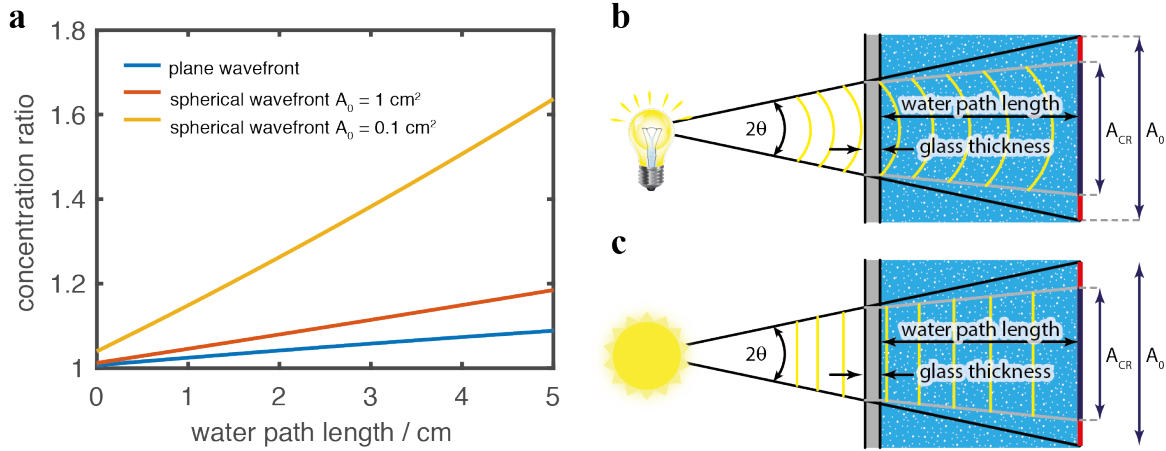

**Fig. S7**

(**a**) Calculated optical concentration ratio of the non-parallel light-beam of solar simulator illumination in PEC cells for plane wavefront and spherical wavefront as a function of water path length. (**b**) Illustration of the spherical wavefront case. The concentration ratio (CR = $A_0/A_{CR}$) depends on the exact sample area $A_0$. (**c**) Illustration of the plane wavefront case. An opening aperture in front of the quartz window of the PEC cell with a diameter of 2 cm was used in this study. The beam divergence was experimentally determined to be $\Theta_V = 1.79\,°$ vertically and $\Theta_H = 2.5\,°$ horizontally.
10

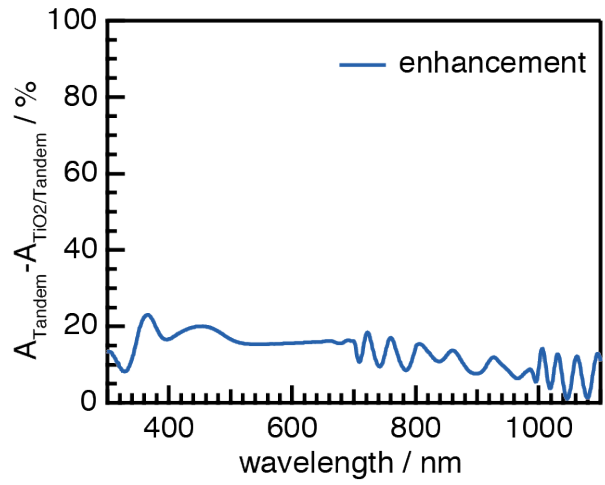

**Fig. S8**

The enhancement of absorption based on the reduction of the reflectivity for the PEC device due to employment of TiO$_2$ layer.

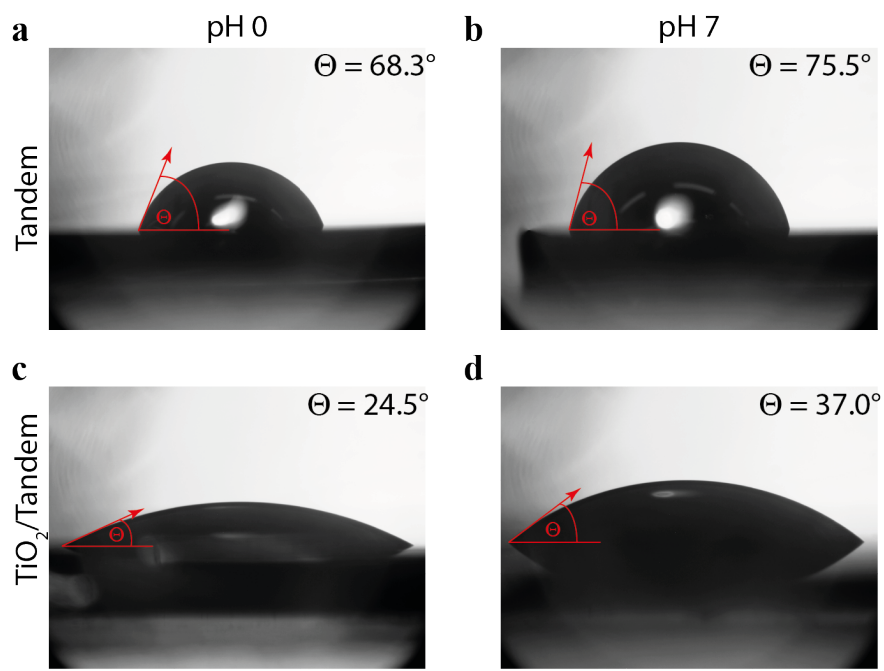

**Fig. S9**

Contact angle measurement for 1 M $HClO_4$ (**a**, **c**) or 0.5 M Phosphate Buffer (**b**, **d**) on the tandem (**a**, **b**) or on the $TiO_2$/tandem (**c**, **d**) sample. The image was analyzed with ImageJ with the help of the 'Drop Analysis' plugin developed at the École polytechnique fédérale de Lausanne (EPFL) [26].

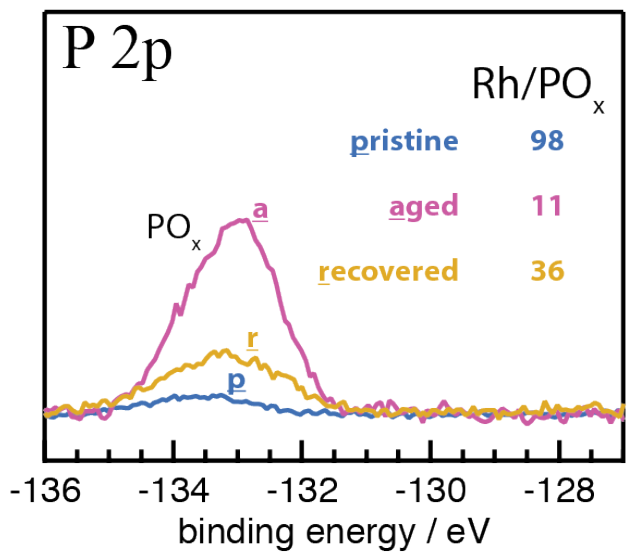

**Fig. S10**

X-ray photoelectron spectra of P 2p core level for the study of Rh catalyst poisoning by $PO_x$ groups in pH 7. Note that no In signal is detected in all samples which indicates the source of phosphate species is the buffer electrolyte rather than tandem corrosion. The ratio of Rh to $PO_x$ is given for pristine sample (p), aged sample (a) and a recovered sample (r). The emersion step reduces the catalyst poisoning and recovers 13.3 % (absolute) of the current loss.

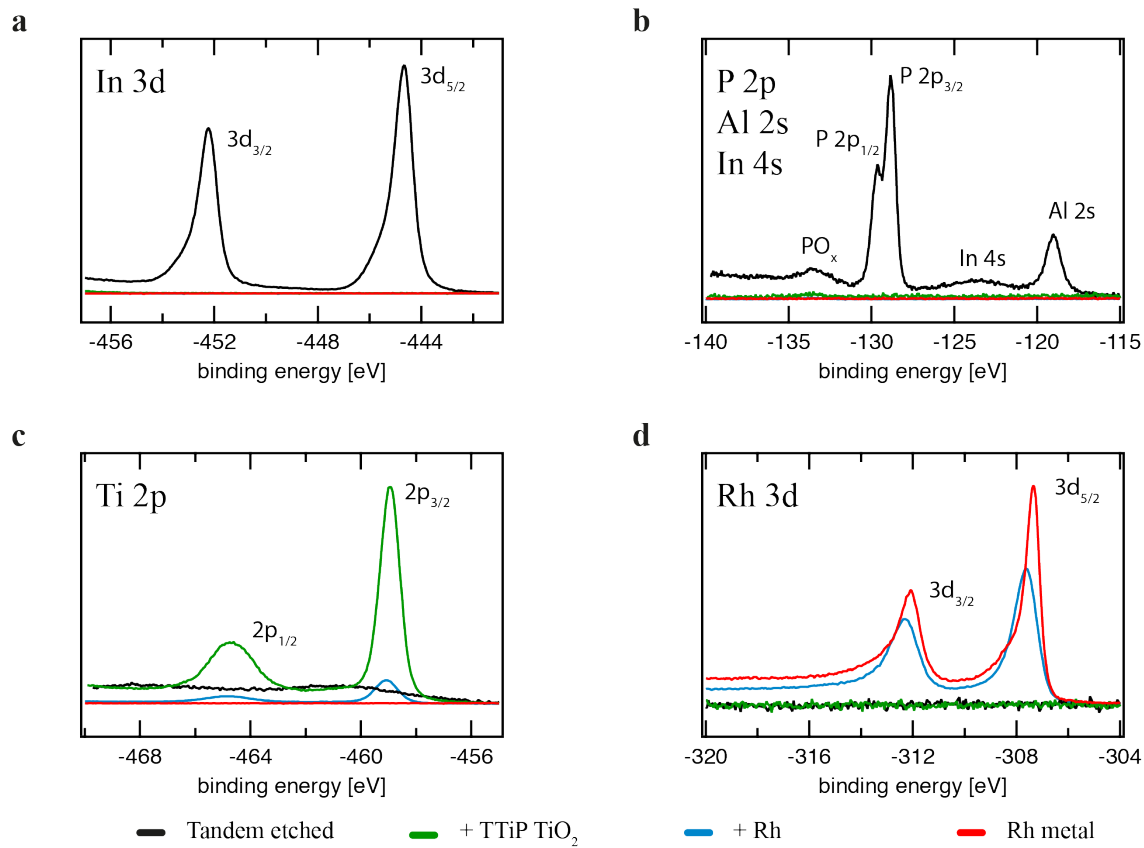

**Fig. S11**

X-ray photoelectron spectra of tandem samples after each step in the PEC device production process: after removing the GaAs/GaInAs cap layer by chemical etching (black curve), after deposition of the TiO$_2$ layer by ALD (green curves); and after photoelectrochemical deposition of Rh nanoparticle catalysts (blue curve). For reference, spectra of metallic Rh electrode are included (red curve). (**a**) In 3d core levels; (**b**) P 2p, In 4s and Al 2s core levels; (**c**) Ti 2p core level; and (**d**) Rh 3d core level.

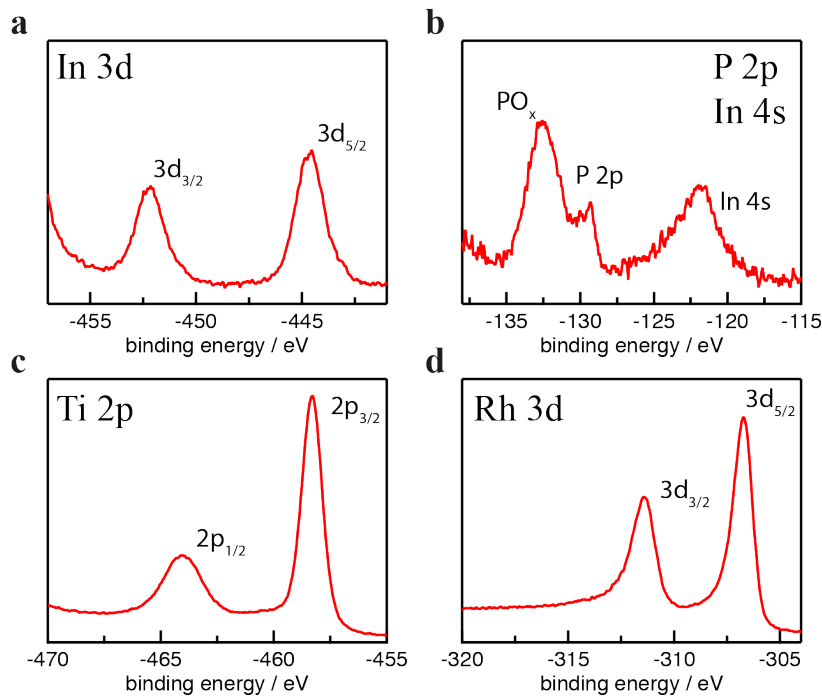

**Fig. S12**

X-ray photoelectron spectra of a Rh/TiO$_2$/Tandem sample after failure in acidic environment: (**a**) In 3d core levels; (**b**) P 2p and In 4s core levels; (**d**) Ti 2p core level; and (**d**) Rh 3d core level. The TiO$_2$ peak enhancement indicates more exposed area upon local detachment of catalyst. However, the maintained prominent Rh peak implies the loss of catalyst is not the limiting factor of degradation. Instead, the appearance of underlying In and PO$_x$ peaks supports the scenario of tandem corrosion due to local TiO$_2$ etching.

**Table S1.**

Reported STH benchmarks from literature with employed bandgaps, achieved STH efficiency, theoretical limit for realistic water splitting ($SQ_{water}$) and ratio of achieved STH to $SQ_{water}$.

|  | **Bandgaps** | **STH %** | **$SQ_{water}$ / %** | **STH to $SQ_{water}$** | **Reference** |
|---|---|---|---|---|---|
| **JCAP/TUI/ISE** | 1.78/1.26 | 19.3 | 22.8 | 0.85 | This work |
| **NREL** | 1.8/1.2 | 16.2 | 24.2 | 0.67 | (11) |
| **TUI/HZB/JCAP/ISE** | 1.78/1.26 | 14 | 22.8 | 0.61 | (5) |
| **JCAP** | 1.84/1.42 | 10.5 | 19.7 | 0.53 | (6) |
| **NREL** | 1.83/1.42 | 10 | 19.7 | 0.51 | (4), (11) |

**Supplementary References**

26. Drop Shape Analysis. *bigwww.epfl.ch* Available at: http://bigwww.epfl.ch/demo/dropanalysis/. (Accessed: 3rd April 2017)